\documentclass[aps,prl,superscriptaddress,twocolumn,epsf,showpacs]{revtex4}
\usepackage{graphicx}

\begin{document}

\title{Controlling growth of nanorod via screening effects}
\author{Da-Jun Shu}
\email{djshu@nju.edu.cn}
\author{Xiang Xiong}
\author{Zhao Wu Wang}
\author{Mu Wang}
\email{muwang@nju.edu.cn}
\author{Ru-Wen Peng}
\author{Nai-Ben Ming }
\affiliation{National Laboratory of Solid State Microstructures and\\
Department of Physics, Nanjing University, Nanjing 210093, China}
\date{\today}

\begin{abstract}
We demonstrate in this report that nanorods can spontaneously grow on the top of a transient mound when the screening effect is considered in island growth. The number of the topmost growing layers is defined as the active length, which decreases as the screening effect is enhanced. It follows that the thickness of the transient mound underneath the nanorod increases as the square of the active length, and a smaller active length favors the formation of nanorod.
\end{abstract}
\maketitle

Due to the size confinement of nanorods, their unique electric and optical properties make them prospective materials for device applications in nanoelectronics and nanophotonics \cite{1Dnano,Au-rod,ZnO-09}. However, controlling the growth of nanorod is usually tricky, and the understanding of the associated growth mechanism remains a challenging task. In metallic growth system, for example, in the absence of catalysts, instead of generating nanorods, very often a structure calld "wedding-cake" or "mound" is generated \cite{mound97,Krug}.   To solve this problem, oblique angle deposition is normally used, in which the upmost surface features shadow  nearby lower surface regions where the incident atoms cannot reach
\cite{Huang08,Huang05,Ye04,Ru-ob}. In addition, in some systems the long-range attractive forces between the incoming atoms and grown structure inherently steer the incident atoms to the upper growth sites, and hence stimulate the growth of nanorods in these regions \cite{wzl}.  The essential feature in these two scenarios is that the topmost structures are superior in capturing incoming atoms. Here we term such a phenomenon as the screening effect. In fact, it is a well established effect in Laplacian growth: Among the protruding growing parts, the front-most tips get more nutrients and hence grow faster.  Although screening effect has been observed to suppress the growth of those being trapped behind and thus form more ordered filament structures \cite{Zhongsh01,DLA83}, the quantitative understanding of the screening effect in the interfacial growth, especially the role of screening effect in keeping the geometrical shape of the nanorods, has not been well addressed so far to the best of our knowledge.

In general, interfacial growth involves landing and accommodation of atoms on the surface, diffusion of adatoms on the surfaces, nucleation, and formation of a new crystalline layer by lateral expansion of the nucleus. The screening effect influences the landing positions of the deposit atoms, and surface diffusion promotes their  redistribution. So both screening effect and surface kinetics govern the distribution of the adatoms on crystal surface, and affect the nucleation and growth processes. The influences of surface kinetics on the growth morphology and size of nanostructures have been extensively studied \cite{Zhang97,Tersoff94,Krug}. Recently we also demonstrated the influence of surface kinetics on the growth mode and radius selection of nanorod growth \cite{shu}.

In this report we focus on the influence of screening effect on the formation of nanorods with uniform cross-section size. Our calculations indicate that the nanorods can spontaneously grow on top of a transient mound when the screening effect has been considered in island growth. The number of the topmost growing layers is defined as the {\it active length}, which decreases as the screening effect is enhanced. Consequently, the thickness of the transient mound underneath the nanorod increases as the square of the active length, and a smaller active length favors the formation of nanorods.

Consider a growing stepped nanostructur with $n$  atomic layers. The dimensionless area of the substrate occupied by the nanostructure is denoted as $A_0$, and that of the $i$-th layer relative to the substrate  is as $A_i$.
 In island growth, the energy barrier at the step is usually larger than that within the terrace. The growth of a partially buried layer is maintained by the  directly deposit atoms in the underneath layer, since the interlayer transport of the adatoms on the buried layers can be negligible.
Assuming adatoms are deposited with rate in the normal direction  of $F$ per surface lattice site.
If there is no screening effect, the deposition is uniform on the surface, and all the layers can keep growing by capturing the deposit atoms. The opposite limiting case is that only the topmost layer is allowed to grow as a result of  extremely strong screening effects. The more realistic situation is that  the moderate strong screening effect leads to a finite number of the active layers, $i.e.$,  the topmost layers which are active in capturing the deposit atoms.  Correspondingly, only the topmost $N_g$  layers  keep grow as the number of atomic layers  $n$ adds up with deposition.  Therefore the quantity $N_g$ can be regarded as a measure of the screening strength: The smaller $N_g$ is, the stronger the screening is. We refer hereafter $N_g$ as the {\it active length}.

 By defining the dimensionless time as $T=Ft$ at deposition time $t$,  the coverage $\theta_i$, which is defined as $A_i/A_0$, increases with time according to the rate equation \cite{Krug,Krug97},
\begin{eqnarray}
&& \frac{d \theta_i}{dT} = 0~~~~~(i\leq n-N_g),  \label{eq3} \\
&& \frac{d \theta_i}{dT} = \theta_{i-1}-\theta_i~~~~~ (n-N_{g}<i<n).  \label{eq1}
\end{eqnarray}
Since there are no sinks for atoms atop, the topmost terrace  absorbs all the atoms landing on layers $n$ and $n-1$, and grows according to
\begin{equation}
\frac{d \theta_n}{dT} = \theta_{n-1}.\label{eq2}
\end{equation}

 \begin{figure}[t]
  \includegraphics[width=7.5cm]{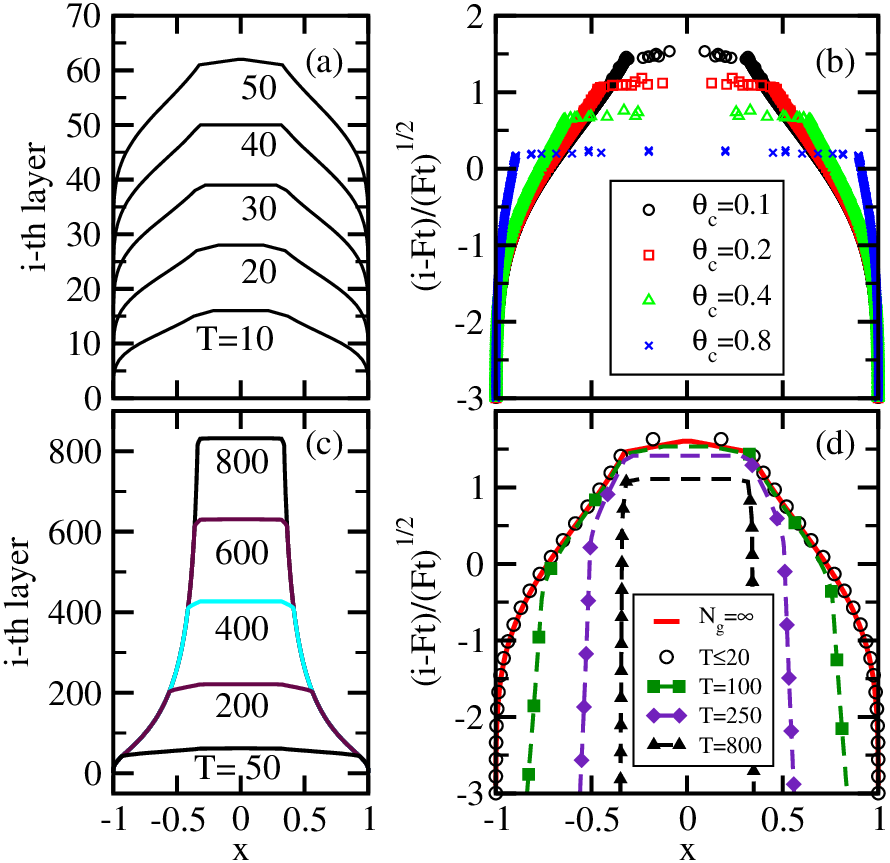}
\caption{Morphologies without screening effect ($N_g = \infty$, upper panels) and $N_g=20$ (lower panels).  (a,c) Surface profiles of morphologies at different  $T$, with characteristic coverage $\theta_c=0.1$. (b) Convergence of the normalized profiles when without screening effect, depending on different $\theta_c$. (d) Normalized profiles with $N_g=20$ and $\theta_c=0.1$ at different $T$.}
\label{fig1}
\end{figure}

The influences of interlayer hoppting rate $\nu'$  and the deposition rate $F$ are included into a  characteristic coverage which is  defined as $\theta_c = A_c/A_0$, where $A_{c} = (7\sqrt{\pi}\nu'/F)^{2/5}$ is the dimensionless characteristic area according to Ref.\ \cite{shu}. Starting from the substrate with area $A_0$ larger than $A_c$, the coverage  of the $i$-th layer $\theta_{i}$ at different time can be obtained  by integrating  Eq.\ (\ref{eq1}) with a time step $\Delta T$. In this work, we use $\Delta T=0.01$, and the influence of different $\Delta T$ is discussed later.

In the general situation of growth, the time interval between subsequent deposition events is much larger than the survival time of the adatom on $A_n$. For the simplest situation where the smallest stable nucleus is a dimer, if a circular cross section is assumed, the nucleation rate on top of $A_{n}$ per unit time can be deduced $\Omega = \frac{F^2A_{n}^{5/2}}{2 \sqrt{\pi} \nu'}$ \cite{shu}. The corresponding nucleation rate  per unit dimensionless time is $\Omega_T= \Omega/F=\frac{7}{2}(A_n/A_c)^{5/2}$.
The random nucleation process is then modeled by introducing a uniform distributed random number between 0 and unit:
Whenever $\Omega_T\Delta T $  is larger than the random number, a new layer is added atop and $n$ increases by one.

 \begin{figure}[t]
 \includegraphics[width=7.5cm]{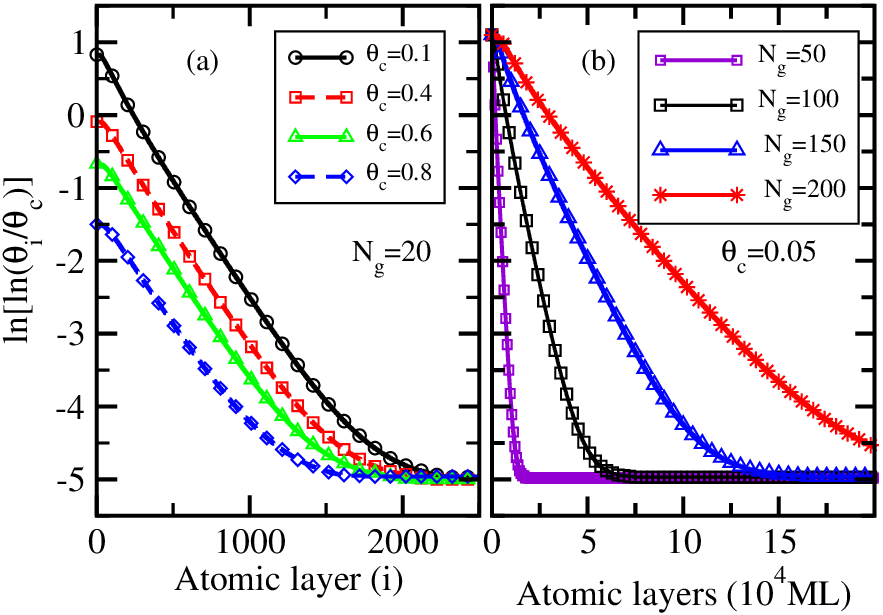}
\caption{(b) Scaled coverage $\theta_i$ after $Ft=1000$ ML at different $\theta_c$. (b) Plots of the scaled coverage of the $i$-th layer $\theta_{i}$ after sufficient deposition for different  $N_g$, while $\theta_c=0.05$.  }
\label{fig2}
\end{figure}

The variation of $\theta_{i}$ is recorded  under specific  growth conditions. The surface profiles without screening effect  at different $T$  are shown in Fig.\ 1(a). The lateral position  on the $x$ axis is  proportional to $ \sqrt{\theta_{i}}$. It shows that the number of the exposed layers increases with time. After re-scaling,   the height profile of the mound converges to a time-independent asymptotic shape which only depends on $\theta_c$,  as shown in  Fig.\ \ref{fig1}(b).  This is a typical characteristic  of the morphology known as {\it wedding-cake} \cite{Krug,Krug97}.

The general evolvement of the surface profile when screening effect is introduced  is shown in  Fig.\ \ref{fig1}(c), where $N_{g}=20$ and $\theta_c=0.1$. It indicates  that the coverage  decreases rapidly as a function of $i$, until it approaches $\theta_c$ within $n_t$ transient layers.  The corresponding re-scaled profiles at different $T$ are given in Fig.\ \ref{fig1}(d). It shows that the curve of  $T\leq N_g$ is superposed on the the one without screening effect, which suggests that the island grows just as in the situation of infinite $N_g$,  keeping the well-known morphology of wedding-cake shape.  Once $n$ is larger than $N_g$, it is obvious that the re-scaled profiles in Fig.~\ref{fig1}(d) gradually depart from the wedding-cake one,  since the bottom $n-N_g$ layers cease growing.

  \begin{figure}[t]
  \includegraphics[width=7.5cm ]{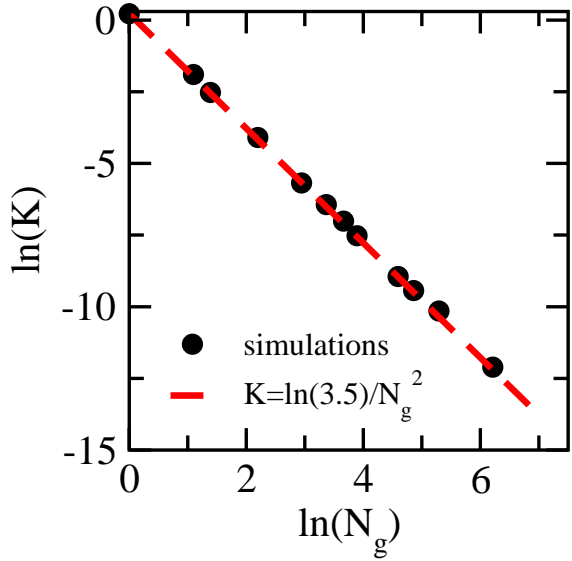}
\caption{Logarithmic plot of the slope of the linear region of the curves shown in Fig.\ \ref{fig2} with $\theta_c=0.01$ (dots) and that of $K= \ln (3.5)/N_g^2$ (dashed line). }
\label{fig3}
\end{figure}

To identify the physical meaning of  $A_c$, we repeat the numerical simulations with different $\theta_c$ while keeping $N_g$ a constant.  For the stable morphologies  obtained after sufficient growth, we find that
$\ln(\ln(\theta_{i}/\theta_c))$ decreases approximately linearly  as a function of $i$, as shown in Fig.\ \ref{fig2}(a),  until it decreases to a  constant $C$ when $i=n_t$, and then it remains. It suggests that a nanorod with area of $A_r=A_c \exp(\exp(C))$ is grown on top of the mound.
The constant $C$ decreases with decreasing  $\Delta T$ used in the simulations.  When $\Delta T=0.01$, $C\simeq -5$ which corresponds to a stable nanorod with area $A_r \simeq 1.00676 A_c$.  It justifies the importance of $A_c$ as a characteristic area \cite{shu}. Realistically, $A_r$ should always be  integral in unit of $a_0^2$,  therefore $A_i/A_c$ decreases and approaches  to $A_r/A_c \simeq  (1+1/A_c)$ as the limit. For the $A_c$ in nanoscale, $C$ is in the range of $-4.5 \sim -6.9$.

In order to study the quantitative  influence of the screening effect, the variation of coverage $\theta_{i}$  for different  $N_g$ is simulated by setting $\theta_c=0.05$. The re-scaled coverage $\ln(\ln(\theta_{i}/\theta_c))$  are shown  in Fig.\ \ref{fig2}(b), as a function of $i$. The main features are similar as that discussed above  for $N_g=20$ in Fig.\ \ref{fig2}(a).   The screening effect just changes the magnitude  of the linear slope:  When $N_g$ is increased as a result of weaker screening effect, the magnitude of the slope becomes smaller correspondingly.

The dependence of slopes on $N_g$ can be obtained by fitting the linear region of the curves in Fig.\ \ref{fig2}(b). The results are plotted logarithmically  in Fig.\ \ref{fig3}, where $K$ is positive denoting the magnitude of the slope. It reveals that $ K$ is inversely proportional to $N_g^2$,
\begin{eqnarray}
K=\ln(3.5)/N_g^2. \label{eqn:K}
\end{eqnarray}
The area of $A_{i}$ is then obtained as, 
\begin{eqnarray}
\frac{A_{i}}{A_c}=\frac{\theta_{i}}{\theta_c}=\left(\frac{1}{\theta_c}\right)^{(\frac{2}{7})^{i/N_g^2}}, (n_1 < i < n_{t})
\label{eqn:Anf}\end{eqnarray}
The pow index becomes $(\frac{2}{7})^i$ when $N_g=1$,  which is consistent with the analytic results in Ref. \cite{shu}.

It is  thus clear that starting from the substrate with cross sectional area $A_0$  larger than $A_c$,
 a nanorod with area of $A_r$ is finally grown on top of a transient mound of $n_t$ layers.  According to Eq.\ (\ref{eqn:Anf}), the number of the transient layers $n_t$ is proportional to $N_g^2$, with the coefficient $a\simeq  [-\ln(\ln(A_r/A_c))-\ln(\ln(A_0/A_c))]/\ln 3.5$.  The value of $a$ is in the range of $2 \sim 7$ for  $\theta_c$ from 0.001 to 0.9 and  $A_c$ in nanoscale.
  The growth of a nanostructure therefore shows different feathers at different duration: Wedding-cake morphologies ($n<N_g$), tapered morphologies ($N_g<n<aN_g^2$) and nanorods with a tapered base ($n>aN_g^2$) can be observed successively.  $A_c$ determines the  area of the approached nanorod, whereas the screening strength determines the transient layers, $n_t=aN_g^2$.  At the limiting case when no screening exists, the well-known wedding cake morphology is obtained.

 In conclusion, we have conducted simulations to study the screening effect on the nanorod growth. We found that  a specific growth system can be well characterized as the characteristic area of $A_{c}$  and the active length $N_g$. $A_c$ determines the  area of the approached nanorod, whereas the screening strength influences the transition process from taper-like morphology to the nanorod, $n_t=aN_g^2$.  Larger screening effect leads to smaller active length $N_g$ and favors nanorod growth.  Therefore  it is possible to control the nanorod growth process by tuning $N_g$ via varying the screening strength, for example by varying the deposition angle in oblique angle deposition.

This work was supported by NSF of China (10974079,10874068 and 10625417) and Jiangsu Province (BK2008012), MOST of China (2010CB630705 and 2006CB921804).


\end{document}